\documentstyle[12pt,aasms4]{article}
\tightenlines

\def    \beq    {\begin{equation}}
\def    \eeq    {\end{equation}}
\def    \ba     {\begin{eqnarray}}
\def    \ea     {\end{eqnarray}}
\def    \H      {{\rm H}}
\def    \HH     {{\rm H}_2}

\def    \K      {{\rm K}}

\def    \bJ     {{\bf J}}

\def    \ba     {{\bf a}}

\def    \cm     {\,{\rm cm}}

\def    \simlt  {\lower.5ex\hbox{$\; \buildrel < \over \sim \;$}}
\def    \simgt  {\lower.5ex\hbox{$\; \buildrel > \over \sim \;$}}
\def    \gtsim  {\simgt}
\def    \ltsim  {\simlt}
\def    \th     {{\rm th}}
\def    \tf     {{\rm tf}}
\def    \trap   {{\rm trap}}

\def\plotBTD#1#2{%
  \expandafter\ifx\csname epsfbox\endcsname\relax
    \immediate\write16{%
        You need to input epsf; I'll do it for you
    }%
    \input epsf
  \fi
  \epsfysize=#2
     \openin 1 #1 \ifeof 1
        \immediate\write16{Can't open #1}%
        \vskip \the\epsfysize
      \else
         \closein 1
         \centerline{\epsfbox{#1}}%
      \fi
}

\lefthead{Lazarian \& Draine}
\righthead{Thermal Flipping}

\begin{document}

\title{Thermal Flipping and Thermal Trapping -- New Elements in Dust Grain Dynamics}

\author{A. Lazarian \&  B.T. Draine}
\affil{Princeton University Observatory, Peyton Hall, Princeton,
NJ 08544}

\begin{abstract}

Since the classical work by Purcell (1979) it has been generally
accepted that most interstellar grains rotate suprathermally.
Suprathermally rotating grains would be nearly perfectly aligned with
the magnetic field by paramagnetic dissipation if not for
``crossovers'', intervals of low angular velocity resulting from
reversals of the torques responsible for suprathermal rotation; during
crossovers grains are susceptible to disalignment by random impulses.

Lazarian and Draine (1997) identified thermal fluctuations within
grain material as an important component of crossover dynamics.  For
$a\gtsim 10^{-5}\cm$ grains, these fluctuations ensure good
correlation of angular momentum before and after crossover resulting
in good alignment, in accord with observations of starlight
polarization.  In the present paper we discuss two new processes which
are important for the dynamics of $a\ltsim10^{-5}\cm$ grains.  The
first -- ``thermal flipping'' -- offers a way for small grains to
bypass the period of greatly reduced angular momentum which would
otherwise take place during a crossover, thereby enhancing the
alignment of small grains.  The second effect -- ``thermal trapping''
-- arises when thermal flipping becomes rapid enough to prevent the
systematic torques from driving the grain to suprathermal rotation.
This effect acts to reduce the alignment of small grains.

The observed variation of grain alignment with grain size would then
result from a combination of the thermal flipping process -- which
suppresses suprathermal rotation of small grains -- and due to H$_2$
formation and starlight -- which drive large grains to suprathermal
rotation rates.

\end{abstract}


\section{Introduction}

One of the essential features of grain dynamics in the diffuse
interstellar medium (ISM) is suprathermal rotation (Purcell 1975,
1979) resulting from systematic torques that act on grains.  Purcell
(1979, henceforth P79) identified three separate systematic torque
mechanisms: inelastic scattering of impinging atoms when gas and grain
temperatures differ, photoelectric emission, and H$_2$ formation on
grain surfaces\footnote{
        Radiative torques suggested in Draine \& Weingartner
        (1996) as a means of suprathermal rotation are effective only 
        for grains
        larger than $5\times 10^{-6}$~cm.
        }. 
The latter was shown to dominate the other two for typical conditions
in the diffuse ISM (P79).  The existence of systematic H$_2$ torques
is expected due to the random distribution over the grain surface of
catalytic sites of H$_2$ formation, since each active site acts as a
minute thruster emitting newly-formed H$_2$ molecules.

The arguments of P79 in favor of suprathermal rotation were so clear
and compelling that other researchers were immediately convinced that
interstellar grains in diffuse clouds should rotate suprathermally.
Purcell's discovery was of immense importance for grain alignment.
Suprathermally rotating grains remain subject to gradual alignment by
paramagnetic dissipation (Davis \& Greenstein 1951), but due to their
large angular momentum are essentially immune to disalignment by
collisions with gas atoms.

Spitzer \& McGlynn (1979, henceforth SM79) showed that suprathermally
rotating grains should be susceptible to disalignment only during
short intervals of slow rotation that they called
``crossovers''\footnote{Crossovers 
	are due to various grain surface processes that change the 
	direction (in body-coordinates) of the systematic torques.
	}. 
Therefore for sufficiently infrequent crossovers suprathermally
rotating grains will be well aligned with the degree of alignment
determined by the time between crossovers, the degree of correlation
of the direction of grain angular momentum before and after a
crossover (SM79), and environmental conditions (e.g., magnetic field
strength $B$).

The original calculations of SM79 obtained only marginal correlation
of angular momentum before and after a crossover, but their analysis
disregarded thermal fluctuations within the grain material.  Lazarian
\& Draine (1997, henceforth LD97), showed that thermal fluctuations
are very important, and result in a high degree of correlation for
grains larger than a critical radius $a_c$, the radius for which the
time for internal dissipation of rotational kinetic energy is equal to
the duration of a crossover.  Assuming that the grain relaxation is
dominated by Barnett relaxation (P79), LD97 found
$a_c\approx1.5\times10^{-5}\cm$, in accord with observations that
indicated that the dividing line between grains that contribute and
those that do not contribute to starlight polarization has
approximately this value (Kim \& Martin 1995).

Here we report the discovery that a new effect of thermal fluctuations
-- which we term {\it ``thermal flipping''} -- should lead to
alignment of even the small grains with $a\ltsim a_c$, {\it if} they
rotate suprathermally.  However, small grains are observed to {\it
not} be aligned.  We argue that this is due to a second effect -- {\it
``thermal trapping''} -- which causes small grains to rotate thermally
a significant fraction of the time, despite systematic torques which
would otherwise drive suprathermal rotation.

In \S2 we review the role of Barnett fluctuations in the crossover
process.  In \S3 we discuss how crossovers influence grain alignment,
and in \S4 we argue that ``thermal trapping'' can account for the
observed lack of alignment of small grains.

\section{Crossovers and Thermal Flipping}

SM79 revealed two basic facts about grain crossovers.  First, they
showed that in the absence of external random torques the direction of
grain angular momentum $\bJ$ remains constant during a crossover,
while the grain itself flips over.  Second, they demonstrated that in
the presence of random torques (arising, for instance, from gaseous
bombardment) the degree of correlation of the direction of $\bJ$
before and after a crossover is determined by $J_{\rm min}$, the
minimum value of $|J|$ during the crossover.  As a grain tends to
rotate about its axis of maximal moment of inertia (henceforth
referred to as ``the axis of major inertia''), SM79 showed that the
systematic torque components perpendicular to this axis are averaged
out and the crossover is due to changes in $J_{\|}$ due to the
component of the torque parallel to the axis.\footnote{
	Indices $\|$ and $\bot$ 
	denote components parallel
        and perpendicular to the axis of major inertia.
        }
Midway through the flipover $J_{\|}=0$ and $J_{\rm min}=J_{\bot}$.

The time scale for Barnett relaxation is much shorter than the time
between crossovers.  For finite grain temperatures thermal
fluctuations deviate the axis of major inertia from $\bf J$ (Lazarian
1994).  These deviations are given by the Boltzmann distribution
(Lazarian \& Roberge 1997) which, for an oblate grain (e.g., an
$a\!\times\! b\!\times\! b$ ``brick'' with $b>a$) is \beq
f(\beta)d\beta= {\rm const}\cdot\sin\beta\exp
\left[-E_k(\beta)/kT_d\right]d\beta~~~;
\label{eq:r1}
\eeq
\beq
E_k(\beta)=\frac{J^2}{2I_z}
\left[1+\sin^2
\beta\left(\frac{I_z}{I_{\bot}}-1\right)
\right]
\label{eq:ek1}
\eeq 
is the kinetic energy, and $\beta$ the angle between the axis of
major inertia and $\bf J$.  We define \beq J_d \equiv
\left(\frac{I_zI_{\bot}kT_d}{I_z-I_{\bot}}\right)^{1/2} \approx (2I_z
kT_d)^{1/2}~~~.
\label{r2}
\eeq 
where the approximation assumes $I_z\approx 1.5I_\perp$, as for
an $a\!\times\!b\!\times\!b$ brick with $b/a=\sqrt{3}$.  The Barnett
relaxation time is (P79) \beq t_{\rm B}=8\times10^7 a_{-5}^7 (J_d/J)^2
\sec ~~~,
\label{t_Bar}
\eeq
where $a_{-5}\equiv a/10^{-5}\cm$.  For $J>J_d$, the most probable
value of $\beta$ for distribution (\ref{eq:r1}) has
$J_\perp=J\sin\beta=J_d$, while for $J<J_d$ the most probable value of
$\beta$ is $\pi/2$.  It follows from (\ref{eq:r1}) that during
suprathermal rotation ($J^2\gg J_d^2$) the fluctuating component of
angular momentum perpendicular to the axis of major inertia $\langle
J_{\bot}^2 \rangle \approx J_d^2$.

SM79 defined the crossover time as $t_c=2
J_{\bot}/|\dot{J_\parallel}|$ where $\dot{J}_\parallel$ is the time
derivative of $J_\parallel$.  If $t_c\ll t_B$, the Barnett
fluctuations can be disregarded during a crossover, and
$J_{\bot}=const\approx J_d$.  The corresponding trajectory is
represented by a dashed line in Fig.~1.  Initially the grain rotates
suprathermally with $\beta\approx 0$; $\beta$ crosses through $\pi/2$
during the crossover and $\beta\rightarrow \pi$ as the grain spins up
again to suprathermal velocities.

The condition $t_c=t_{\rm B}$ was used in LD97 to obtain a critical
grain size $a_{\rm c}\approx 1.5\times 10^{-5}$~cm.  It was shown that
$t_c < t_B$ for $a>a_{\rm c}$, and paramagnetic dissipation can
achieve an alignment of $\sim 80\%$ for typical values of the
interstellar magnetic field.  If paramagnetic aligment were suppressed
for $a<a_{\rm c}$ this would explain the observed dichotomy in grain
alignment: large grains are aligned, while small are not.

What spoils this nice picture is that sufficiently strong thermal
fluctuations can enable crossovers: fluctuations in $\beta$ span
$[0,\pi]$ and therefore have a finite probability to flip a grain over
for an arbitrary value of $J$.  The probability of such fluctuations
is small for $J^2\gg J_d^2$, but becomes substantial when $|J|$
approaches $J_d$.  Indeed, it is obvious from (\ref{eq:r1}) that in
the latter case the probability of $\beta\sim \pi/2$ becomes
appreciable.  LD98 show that the probability per unit time of a
flipover due to fluctuations is 
\beq t_\tf^{-1}\approx t_{\rm B}^{-1}
\exp\left\{(1/2)\left[(J/J_d)^2-1\right]\right\} ~~~.  
\eeq 
Whether the grain trajectory is approximately a straight line in
Fig.~1, (a ``regular crossover''), or two lines connected by an arc (a
``thermal flip'') depends on the efficacy of the Barnett relaxation.
Roughly speaking, thermal flipping happens when $t_\tf$ \ltsim
$J/\dot{J}$.  If $J\approx J_d$ the ratio of the flipping and
crossover time $t_\tf/t_c \approx t_{\rm B}/t_c$.  The latter ratio
was found in LD97 to be equal to $(a/a_c)^{13/2}$.  Therefore flipping
was correctly disregarded in LD97 where only grains larger than $a_c$
were considered, but should be accounted for if $a<a_c$.

The last issue is the problem of multiple flips: a grain with $\beta >
\pi/2$ can flip back.  Thermal flips do not change $\bJ$.  Therefore
after a flip (from quadrant $\beta<\pi/2$ to quadrant $\beta>\pi/2$ in
Fig.~1) the grain has the same $\bJ$ as before the flip.  For $J>J_d$,
the thermal distribution (\ref{eq:r1}) has two most probable values of
$\beta$: $\beta_-=\sin^{-1}(J_d/J)$, and $\beta_+=\pi- \beta_+$.  For
both $\beta_-$ and $\beta_+$ we have $J_\perp=J_d$.  If we idealize
the grain dynamics as consisting of systematic torques changing
$J_\parallel$ with $J_\perp=const$, plus the possibility of
instantaneous ``flips'' (at constant $\bf J$) between $\beta_-$ and
$\beta_+$, then we can estimate the probability of one or more
``flips'' taking place during a crossover.  Let
$\phi(\beta)d\beta=t_\tf^{-1}dt$ be the probability of a flip from
$\beta$ to $\pi-\beta$ while traversing $d\beta$.  The probability of
zero flips between $0$ and $\beta$ is
$f_{00}(\beta)=\exp[-\int_0^\beta \phi(x^\prime)dx^\prime]$.  The
probability of a ``regular crossover'' (zero flips) is
$p_{00}=f_{00}(\pi) =e^{-2\alpha}$, where 
\beq
\alpha\equiv\int_0^{\pi/2}\phi(x) dx \approx
\sqrt{\frac{\pi}{2}}\frac{t_c}{t_B} =
\sqrt{\frac{\pi}{2}}\left(\frac{a_c}{a}\right)^{13/2} ~~~.  
\eeq

Similarly, $df_{10}= f_{00}^2 \phi d\beta$ is the probability of one
forward flip in the interval $d\beta$, with no prior or subsequent
flips, and the probability of exactly one forward and zero backward
flips during the crossover is
$p_{10}=f_{10}(\pi/2)=(1-e^{-2\alpha})/2$.  Therefore the probability
of one or more backward flips is $1-p_{00}-p_{10}=
(1-e^{-2\alpha})/2\rightarrow 1/2$ for $a\ll a_{\rm cr}$.

\section{Efficacy of Paramagnetic Alignment}

SM79 showed that disalignment of suprathermally rotating grains occurs
only during crossovers whereas thermally rotating grains undergo
randomization all the time.  Consider a grain subject to random
(nonsystematic) torques which provide an excitation rate $\Delta
J^2/\Delta t$, and damping torque $-{\bf J}/t_d$, where $t_d$ is the
rotational damping time.  Thermally rotating grains have $\langle
J^2\rangle = (1/2)t_d(\Delta J^2/\Delta t) \equiv J_{th}^2$.  This
definition of thermally rotating grains encompasses grains whose
rotation is excited by {\it random} H$_2$ formation, cosmic ray
bombardment etc.\ -- so long as the associated torques have no
systematic component.  For suprathermally rotating grains $J^2\gg
J_{th}^2$.

In what follows we roughly estimate the efficacy of grain alignment
for $t_c \gg t_{\rm B}$, i.e., $a < a_c$.  Following P79 we assume
that H$_2$ torques are the dominant spin-up mechanism.

A crossover requires $N\sim J_{\rm min}/\langle \Delta J_z \rangle$
impulse events, where the mean impulse per recombination event (see
SM79) $\langle \Delta J_z \rangle \approx \left(2 m_{\rm H} a^2
E/3\nu\right)^{1/2}$ where $\nu$ is the number of active recombination
sites, $E$ is the kinetic energy per newly-formed H$_2$, and $J_{\rm
min}$ is the minimum $J$ reached during the crossover.  If $N$ is
multiplied by the sum of mean squared random angular momentum impulses
$(\langle \Delta J_z^2 \rangle + \langle \Delta J^2_{\bot} \rangle)$
it gives the mean squared change of $J$ during a crossover.  Therefore
the mean squared change of angle during a flipping-assisted crossover
is
\beq
\label{eq:<F>}
\langle F\rangle \approx 
\frac{N(\langle\Delta J_z^2 \rangle +\langle \Delta J^2_{\bot} \rangle)}
{J_{\rm min}^2} \approx
\frac{(\langle\Delta J_z^2 \rangle +
\langle \Delta J^2_{\bot} \rangle)}{J_{\rm min} \langle \Delta J_z \rangle}
\eeq
which differs only by a factor of order unity from the expression for
disorientation parameter $F$ in SM79, provided that $J_{\rm min}$ is
used instead of $J_{\rm \perp}$.  The latter is the major difference
between the regular crossovers that were described by SM79 and LD97
and our present study.  SM79 and LD97 dealt with the case for which
flipping is negligible and the disorientation was mostly happening
when $J_{\parallel} \rightarrow 0$ due to the action of regular
torques, in which case $J_{\rm min}\approx J_d$.  As flipping becomes
important, $J_{\rm min} > J_d$ is obtained from the condition
$t_\tf(J_{\rm min})\approx J_{\rm min}/\dot{J}$.

Grain alignment is measured by $\sigma\equiv
(3/2)(\langle\cos^2\theta\rangle -1/3)$ where $\theta$ is the angle
between the magnetic field direction and $\bf J$.  Generalizing LD97,
\beq
\label{eq:sigma}
\sigma \approx A\left[1+\frac{3}{\delta_{\rm eff}}\left(
\frac{{\rm arctan}\sqrt{e^{\delta_{\rm eff}}-1}}
     {\sqrt{e^{\delta_{\rm eff}}-1}}-1\right)\right]
	+ (1-A)\sigma_\th ~,
\label{eq:sig1}
\eeq
\beq
\label{eq:delta_eff}
\delta_{\rm eff}
=
\frac{2 C \bar{t}_b/t_r}{\left[1-\exp(-\langle F\rangle)\right]}
\approx
\frac{2.6 C t_d/t_r}{\left[1-\exp(-\langle F \rangle)\right]}
\left(1+\frac{t_L}{t_d}\right)
\label{eq:x-fact}
\eeq 
where $A=C=1$ in LD97 theory, $\sigma_\th$ is the alignment
parameter for thermally rotating grains (Roberge \& Lazarian 1998),
$t_r$ is the paramagnetic damping time (Davis \& Greenstein 1951),
$\bar{t}_b\approx 1.3 (t_d+t_L)$ is the mean time back to the last
crossover (P79), and $t_L$ is the resurfacing time.  For typical ISM
conditions ($n_{\rm H}=20$~cm$^{-3}$, $B=5\mu$G) $t_d/t_r\approx 0.05
/ a_{-5}$.

Expressions (\ref{eq:sig1}) and (\ref{eq:x-fact}) (with $A=C=1$) were
obtained in LD97 assuming that grains spend nearly all their time
rotating suprathermally, except for brief crossover events with a
characteristic disorientation parameter $\langle F\rangle$.  We now
argue that a significant fraction of the small grains do {\it not}
rotate suprathermally, and an appreciable fraction of crossovers have
$\langle F\rangle\gtsim 1$.

\section{Thermal trapping of small grains}

P79 theory of suprathermal rotation did not take into account the
``thermal flipping'' process discussed here.  We now argue that
thermal flipping will suppress the suprathermal rotation of very small
grains.

With the Barnett relaxation time $t_{\rm B}$ from (\ref{t_Bar}), the
ratio $t_{\rm B}/t_d\approx 1\times10^{-4} (15/T_d) (J_d/J)^2
a_{-5}^6$, where the drag time $t_d$ is evaluated for a diffuse cloud
with $n_\H=30\cm^{-3}$ and $T=100\K$.  Thus the timescale for a
thermal flip is
\beq
t_\tf/t_d
\approx 10^{-4}a_{-5}^6
(J_d/J)^2
\exp\left\{(1/2)\left[(J/J_d)^2-1\right]\right\}
\label{eq:tf1}
\eeq
showing that thermal flipping is strongly favored for small grains.
The critical question now is: Can the systematic torques drive small
grains to large enough $J$ to suppress the thermal flipping, or is the
thermal flipping sufficiently rapid to suppress the superthermality?

Consider a grain with a systematic torque $(G-1)^{1/2}J_\th/t_d$ along
the major axis (fixed in grain coordinates).  The condition $J_{\rm
min}=\dot{J}\cdot t_\tf(J_{\rm min})$ becomes
\beq
\label{eq:jmin}
t_\tf/t_d = (J_{\rm min}/J_{\rm th})/(G-1)^{1/2} ~~~.
\eeq
Thermal flipping causes the systematic torque to randomly
change sign in inertial coordinates, so that
\beq
\langle J^2\rangle = J_\th^2 +(G-1) J_\th^2 t_\tf/(t_\tf+t_d) ~~~,
\eeq
giving a condition for $t_\tf$ in terms of $\langle J^2\rangle$:
\beq
t_\tf/t_d = 
\left[\langle (J/J_\th)^2\rangle-1\right]/
\left[G-\langle(J/J_\th)^2\rangle\right] ~~~.
\label{eq:tf2}
\eeq
For given $a$, $G$, and $(J_{th}/J_d)^2$, (\ref{eq:tf1}) and
(\ref{eq:tf2}) have either one or three solutions for $\langle
J^2\rangle$.  If $\langle J^2\rangle^{1/2}$ has multiple solutions
$J_1\!<\! J_2\! <\! J_3$, the intermediate solution $J_2$ is unstable:
if $J_1 \!<\! J\! <\! J_2$, then $t_\tf$ from (\ref{eq:tf1}) is
smaller than the value required by (\ref{eq:tf2}), so $J\rightarrow
J_1$; if $J_2 \!<\! J \!<\! J_3$, then $J\rightarrow J_3$.  In the
former case thermal flipping leads to suppression of suprathermal
rotation: if the grain enters the region $J < J_2$, then it is trapped
with $J\approx J_1$ until a fluctuation brings it to $J > J_2$.  The
timescale for such a fluctuation is $t_{\rm trap}\approx t_d
\exp[(J_2/J_\th)^2]$.  We refer to this phenomenon as {\it ``thermal
trapping''}.

As an example, consider $(J_{\rm th}/J_{\rm d})^2 = 5$, $G=10^3$ and
$a_{-5}=0.5$.  Thermal flipping takes place during a crossover at
$J_{\rm min}^2\approx5.9J_\th^2$.  If $J^2$ drops below
$J_2^2=4.5J_\th^2$, the grain will be thermally trapped.  For this
case thermal flipping will tend to maintain the grain at $\langle
J^2\rangle \approx J_1^2\approx 1.02(J_\th)^2$, unless thermal
fluctuations succeed in getting the grain to $J>J_2$, in which case
thermal flipping is unable to prevent the grain spinning up to a
superthermality $(J/J_\th)^2\approx G=10^3$.  For this example the
thermal trapping time $t_\trap\approx 50 t_d$.  During this time
paramagnetic alignment will be minimal.  Grains that escape the
thermal trap become suprathermal and align on the timescale for
paramagnetic alignment.

Let $\eta$ be the probability per crossover of becoming ``thermally
trapped''.  The fraction of grains which are not thermally trapped at
any time is $x=\bar{t}_b/(\bar{t}_b+\eta t_\trap)$.  We can estimate
the grain alignment by using (\ref{eq:sigma}) and (\ref{eq:delta_eff})
with $A=x$ and $C=[1-\exp(-\langle F\rangle)]\{\eta+(1-\eta)
[1-\exp(-\langle F\rangle)]\}^{-1}$.

During a crossover, the first thermal flip takes place at $J\approx
J_{\rm min}$, only a bit larger than $J_2$, the thermal trapping
boundary.  We have seen above that for $a < a_c$, $\sim50\%$ of
crossovers involve one or more ``backward'' flips.  We do not know
what fraction of the crossovers end up ``thermally trapped'', but we
speculate that it could be appreciable, say $\eta\sim 0.1$.

The time between crossovers is of the order of the damping time $t_d$
(see P79).  Returning to our example of a grain with $a_{-5}=0.5$, for
which we estimated $t_\trap\approx 50t_d$, we see that the fraction of
grains which are not trapped $x=1/(1+50\eta t_d/\bar{t}_b)$ could be
small if $\eta \gtsim 0.1$.  More detailed studies of the dynamics
(Lazarian \& Draine 1999) will be required to estimate $\eta$, and to
provide more reliable estimates of $t_\trap$, before we can
quantitatively estimate the degree to which thermal trapping will
suppress the alignment of small grains.

Inspection of Fig. 2 shows that thermal trapping solutions are only
found if $G$ is not too large (e.g. for $G=10^5$ we have no thermal
trapping solution in Fig. 2 for $a_{-5}=0.5$).

Such degrees of suprathermality would follow from variations of
accomodation coefficient and photoelectric yield, but higher values
were obtained in the literature for the case of H$_2$ torques (P79,
LD97).  For example, LD97 estimate $G=2\times 10^7
a_{-5}(\gamma/0.2)^2(10^{11}\cm^{-2}/\alpha)$, where $\alpha$ is the
surface density of active recombination sites.  The values of $G
\ltsim 10^4$ required for thermal trapping to be possible for
$a_{-5}\approx 0.5$ grains would appear to require
$(\gamma/0.2)^2/\alpha \ltsim 10^{-14}\cm^2$.  If essentially every
surface site is an active chemisorption site, we could have
$\alpha\approx 10^{15}\cm^{-2}$; alternatively, it is conceivable that
$\gamma\ll 1$ for very small grains.  The latter idea was advocated by
Lazarian (1995), who found that oxygen poisoning of catalytic sites is
exponentially enhanced for grains with $a<10^{-5}$~cm.  Recent
experimental work (Pironello et al.\ 1997a,b) suggests that $\gamma$
may be much smaller than is usually assumed.  Moreover, recent
research (Lazarian \& Efroimsky 1998, Lazarian \& Draine 1999) shows
that faster processes of internal relaxation are possible. These
processes should enable thermal trapping for larger values of $G$.

\section{Conclusions}

We have found that ``thermal flipping'' is a critical element of the
dynamics of small ($a\ltsim 10^{-5}$~cm) grains.  If small grains
rotate suprathermally, then thermal flipping would promote their
alignment by suppressing disalignment during ``flipping-assisted''
crossovers.  Since small grains are observed to not be aligned, it
follows that most must not rotate suprathermally.

One way for small grains to not rotate suprathermally would be for the
systematic torques from $\HH$ formation and photoelectric emission to
be much smaller than current estimates.  However, we also find that
thermal flipping can result in ``thermal trapping'', whereby rapid
thermal flipping can prevent systematic torques from driving small
grains to suprathermal rotation rates.  As a result, at any given time
an appreciable fraction of small grains are thermally trapped and
being disaligned by random processes.

The thermal trapping effect is of increasing importance for smaller
grains, and may explain the observed minimal alignment of $a\ltsim
5\times10^{-6}\cm$ dust grains (Kim \& Martin 1995).

\acknowledgements

This work was supported in part by NASA grants NAG5-2858 and NAG5-7030,
and NSF grant AST-9619429.

\bigskip\plotBTD{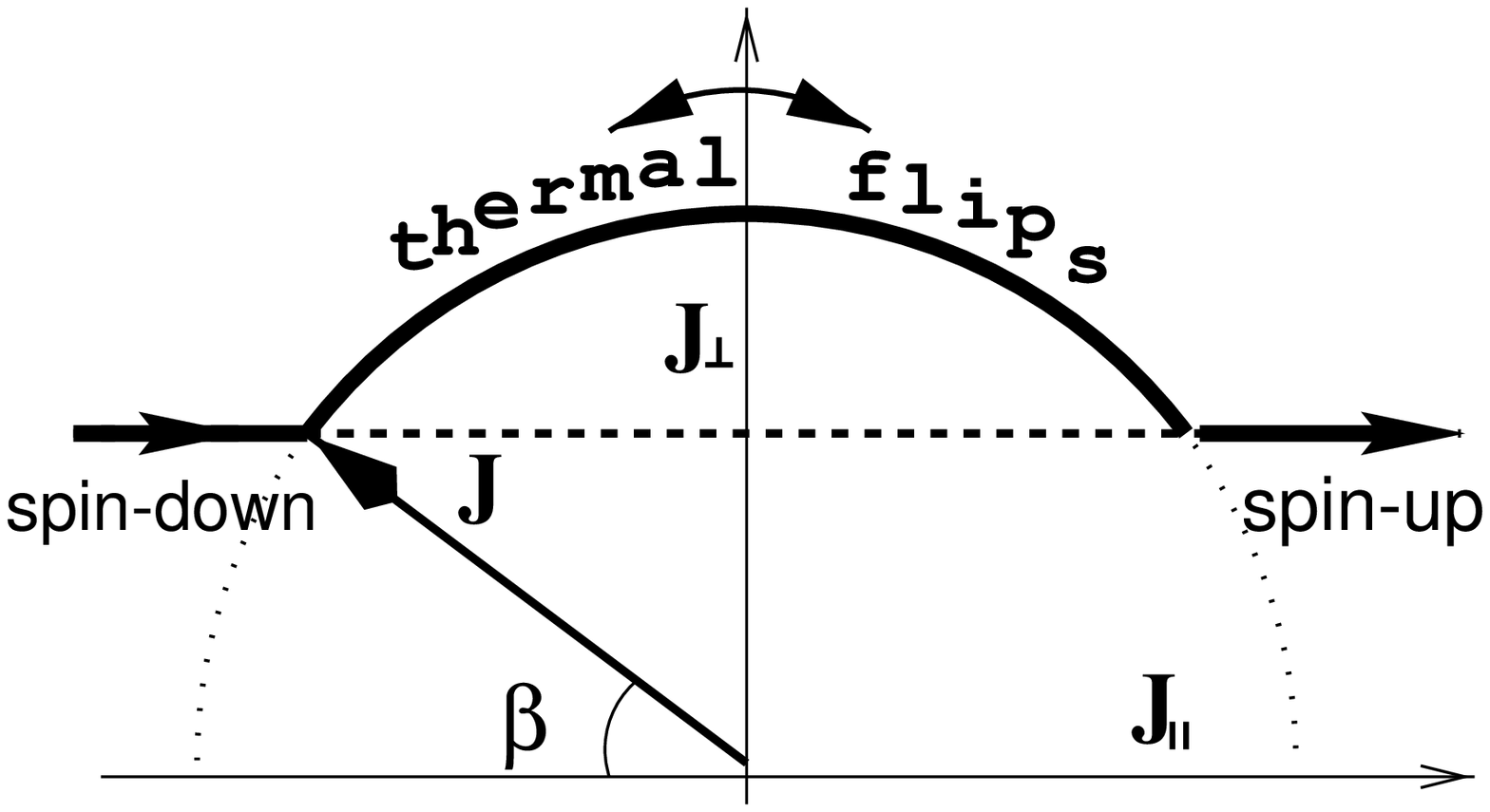}{8.0cm}
\figcaption[f1.eps]{
        Grain trajectory on the $J_\perp$ -- $J_\parallel$ plane, where
        $J_\perp$ and $J_\parallel$ are components of $\bJ$ 
        perpendicular or parallel
        to the grain's principal axis of largest moment of inertia.
        The solid trajectory shows a ``thermal flip'', while the broken line
        shows the ``regular'' crossover which would occur in the absence of
        a thermal flip.
        }
\plotBTD{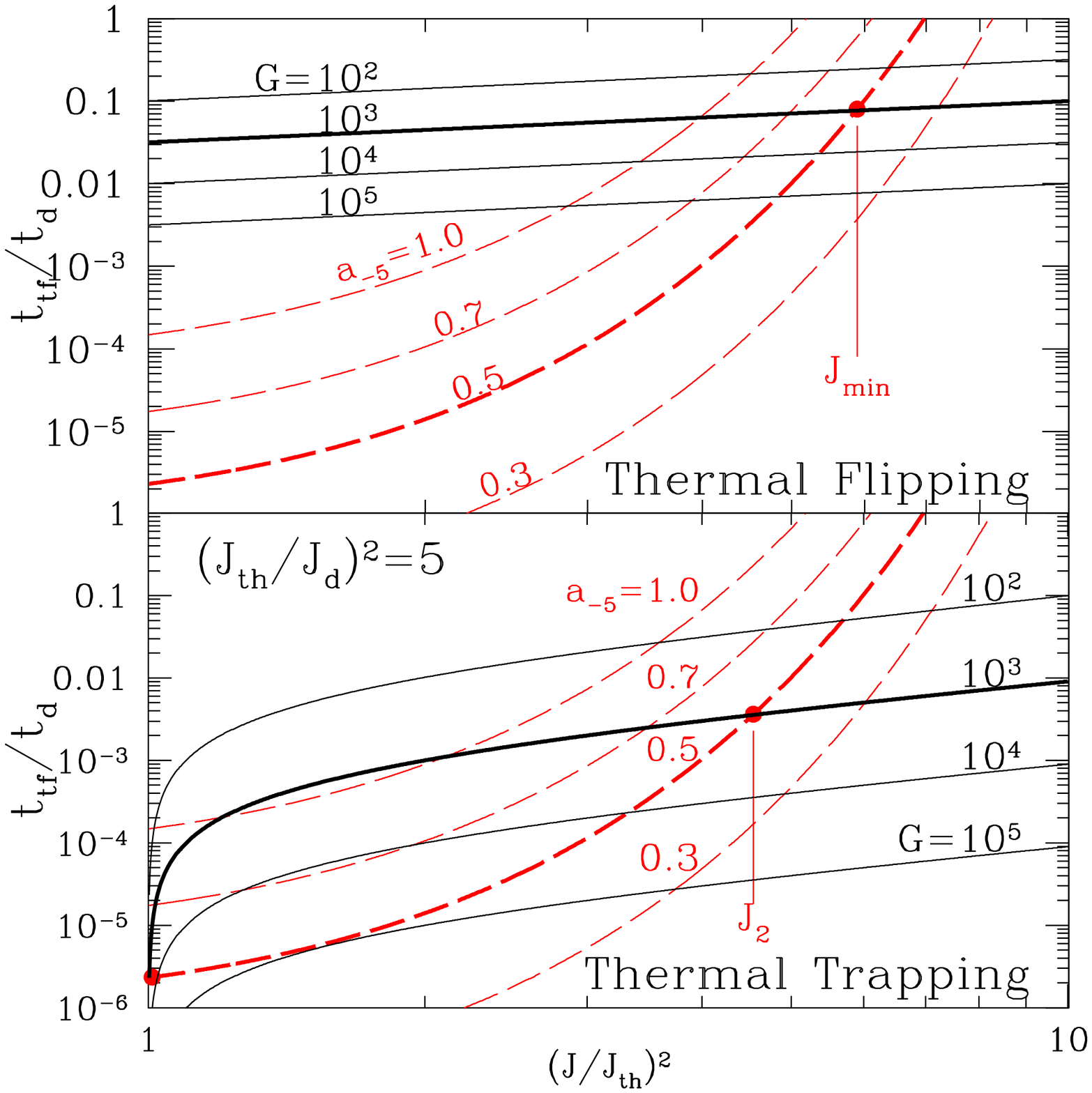}{15cm}
\figcaption[f2.eps]{
        \label{fig:2}
	Upper panel: Eq.\ (\protect{\ref{eq:tf1}}) 
	[broken lines, labelled by $a_{-5}$] and
	eq.\ (\protect{\ref{eq:jmin}}) [solid lines, labelled by $G$].
	Dot shows $J_{\rm min}$ for flipping-assisted crossover of $a_{-5}=0.5$
	grain with $G=10^3$.
	Lower panel: Eq.\ (\protect{\ref{eq:tf1}}) [broken lines,
	labelled by $a_{-5}$], and
	eq.\ (\protect{\ref{eq:tf2}}) 
	[solid lines, labelled by $G$].
	Dots show ``equilibrium'' solutions $J_1$ and $J_2$
	for $a_{-5}=0.5$ and $G=10^3$ (see text).
	}

\begin{thebibliography}{}
\bibitem[]{DG51} Davis, J., \& Greenstein, J.L. 1951, \apj, 114, 206.
\bibitem[]{DW96} Draine, B.T., \& Weingartner, J.C. 1996 ApJ, 470, 551.
\bibitem[]{KM95} Kim, S.-H., \& Martin, P. G. 1995, ApJ, 444, 293 
\bibitem[]{La94} Lazarian, A. 1994, MNRAS, 268, 713
\bibitem[]{La95a} Lazarian, A. 1995, MNRAS, 274, 679
\bibitem[]{LD97} Lazarian, A., Draine, B.T. 1997, ApJ, 487, 248
\bibitem[]{LD98} Lazarian, A., \& Draine B.T. 1999, in preparation
\bibitem[]{LE98} Lazarian, A., \& Efroimsky, M. 1998, MNRAS, in press
\bibitem[]{LR96} Lazarian, A., \& Roberge, W.G. 1997 \apj, 484, 230
\bibitem[]{PLSV97} Pirronello, V., Liu, C., Shen, L., \& Vidali, G. 1997a,
	ApJ, 475, L69
\bibitem[]{PBLSV97} Pirronello, V., Biham, O., Liu, C., Shen, L., \& Vidali, G.
	1997b, ApJ, 483, L131
\bibitem[]{Pu75} Purcell, E.M. 1975, 
        in {\it The Dusty Universe}, eds~G.B.~Field \&
        A.G.W.~Cameron, New York, Neal Watson, p.~155.
\bibitem[]{Pu79} Purcell, E.M. 1979, ApJ, 231, 404.
\bibitem[]{RL98} Roberge, W.G., \& Lazarian, A. 1998, MNRAS, submitted
\bibitem[]{SM79} Spitzer, L. Jr, \& McGlynn, T.A. 1979, \apj, 231, 417.

\end{thebibliography}
\end{document}